\documentstyle[12pt]{article}
\newcount\who 
\input psfig.tex

\begin{document}
\pagestyle{empty}

\null
\vskip 1.0 truecm
{\baselineskip=22truept\parindent=0pt
{\vbox{\vskip 1.0 truecm\rightskip=0pt plus1fill
{\Large\bf Stability of cosmological \break detonation fronts} 
\leftskip=0pt plus1fill }} } 

\bigskip
\bigskip
\bigskip
\medskip
\centerline{ \bf{Luciano Rezzolla}} \par
\bigskip
\medskip
\centerline{{\it Scuola Internazionale di Studi 
Avanzati, Trieste, Italy} }\par
\bigskip
\centerline{\tenrm Electronic address:
{\tt rezzolla@neumann.sissa.it} }\par 
\bigskip 
\medskip 
\begin{abstract} 
\bigskip 
\noindent 
 We present results of a linear stability analysis of relativistic
detonation fronts, which have been considered as representing phase
interfaces in cosmological first order phase transitions. After discussing
general stability conditions for detonation fronts, we concentrate on the
properties of the fronts with respect to corrugation instabilities and
discuss separately the cases of Chapman-Jouguet and strong detonation
waves. Contrarily to what recently claimed, we find that strong
detonations are both evolutionary and stable with respect to corrugations
of the front. Moreover, Chapman-Jouguet detonations appear to be
unconditionally linearly stable. The implications of the stability results
for first order cosmological phase transitions are presented and a
discussion of the causal structure of reaction fronts is also given. 
\end{abstract} 
\vskip 1.0 truecm 
\bigskip{PACS number(s): 98.80.Cq, 95.30.Lz}
 
\vfill\eject

\setcounter{page}{1}
\pagestyle{plain}
\vsize=20.0truecm 
\hsize=15.0truecm 
\voffset=1.0truecm 
\hoffset=-0.75truecm
\baselineskip=18pt plus 2pt minus 1pt
\parindent=1.0truecm

{\who=1\bigskip\medskip\bigskip\goodbreak
{\Large\bf\noindent\hbox{I.}\hskip 0.5truecm Introduction}\nobreak
\bigskip\medskip\nobreak\who=0}

	Reaction fronts within cosmological scenarios have been the
subject of extensive recent investigation. In the classical treatment they
are described as moving surfaces by means of which a suitable gas mixture
undergoes a chemical transformation with liberation of heat and with the
gas being either accelerated and decompressed or decelerated and
compressed as it passes through the front. The study of the microphysics
in the narrow region where the reaction processes take place is extremely
complicated and an exhaustive theory of it within the present context has
not been reached yet. Nevertheless, a satisfactory hydrodynamical
description of reaction fronts can be achieved when these are treated as
discontinuity surfaces of infinitesimal and constant width across which
rapid changes in the fluid variables occur. This approximation is
certainly good if the front has a thickness which is much smaller than the
typical length scale for the variation of the flow variables and if the
thermal and the viscous time scales are much smaller than the one set by
the motion of the front (we shall assume that these requirements are met
within the scenarios of cosmological phase transitions considered here).
In this respect, reaction fronts are very similar to the better known
shock fronts (with which they share some properties) and can be described
by means of the same mathematical theory \cite{cf76,ll93,bl82,fd79,f78}. 

	Reaction fronts can occur in nature in connection with a variety
of different classes of phenomena and in general they can be distinguished
into {\it deflagrations} and {\it detonations} according to whether the
front is subsonic or supersonic relative to the medium ahead of the
\hbox{front.}\footnote{The flow regions ahead
		of and behind a propagating front will be here also 
		referred to as the ``upstream'' and the ``downstream'' 
		regions respectively.}
Deflagrations (and detonations) can be further classified as {\it
weak} or {\it strong} according to whether the velocity of the medium
behind is subsonic (supersonic) or supersonic
(subsonic)\footnotesep 15pt\footnote{Unless specified, all the 
		velocities are meant to be referred to the front 
		rest frame.}. 
		As a general rule, the fluid velocity entering a
deflagration front is always smaller than the fluid velocity going out of
it, while the opposite is true for a detonation front. An additional and
special class of reaction fronts is the one for which the velocity of the
fluid out of the front is exactly equal to the local sound speed. These
fronts are called {\it Chapman-Jouguet} \hbox{deflagrations/detonations},
and represent a specifically interesting class of phenomena. (The
classification of the various reaction fronts is summarized in \hbox{Table
I,} where $v_1$ is the fluid velocity ahead of the front, $v_2$ the fluid
velocity behind the front and $c_{s1}$, $c_{s2}$ are the sound speeds on
either side). 

	It can be shown that Chapman-Jouguet processes yield stationary
values both for the front velocity relative to the fluid ahead and for
entropy of the fluid which has gone across the front \cite{cf76}. More
precisely, $v_1$ and the entropy of the fluid behind the front are at a
maximum in the case of a {\it Chapman-Jouguet deflagration}, while $v_1$
and the entropy of the fluid behind are at a minimum in the case of a {\it
Chapman-Jouguet detonation}, which is then the slowest of all possible
detonations. Chapman-Jouguet detonations represent a particularly relevant
class of reaction fronts, for which the speed of the front is completely
determined in terms of the boundary conditions and of the energy--momentum
conservation. This privileged nature is furthermore underlined in the
``Chapman-Jouguet hypothesis'', according to which detonations in chemical
burning should occur {\it only} under the form of Chapman-Jouguet
detonations.  A proof of this can be found in \cite{cf76,ll93} but it is
important to stress that because of the differences between chemical
combustion and phase transitions, the validity of the Chapman-Jouguet
hypothesis cannot be extended to the context of cosmological phase
transitions \cite{l94}. 
		
	Although in nature detonations in chemical burning often appear
accompanied by nonlinear effects (such as transverse shock waves or
turbulence \cite{fd79}), it seems that the hypothesis is generally
verified to a good approximation, with detonation fronts which although
complicated, propagate at a constant velocity close to the theoretical
Chapman-Jouguet value \cite{lve61,w65,l88}. 

\begin{table}
\begin{tabular}{c l|c|c|}
$\hskip 0.8 truecm $ & $\ $ &
$\ \; {\rm DEFLAGRATIONS} \ $ & 
$\ \; {\rm DETONATIONS} \ $  
\\
$\hskip 0.8 truecm $ & $\ $ &
$\ \; (v_1 < v_2) \ $ & 
$\ \; (v_1 > v_2) \ $  
\\ [1truemm] \cline{2-4} 
& & & \\[1truemm] 
$\hskip 0.8 truecm $ & 
{\rm Weak} & 
$v_1 < c_{s1},\;\;v_2 < c_{s2}$ & 
$v_1 > c_{s1},\;\;v_2 > c_{s2}$ 
\\[2truemm] 
$\hskip 0.8 truecm $ & 
{\rm Chapman--Jouguet} & 
$v_1 < c_{s1},\;\;v_2 = c_{s2}$ & 
$v_1 > c_{s1},\;\;v_2 = c_{s2}$ 
\\[2truemm] 
$\hskip 0.8 truecm $ & 
{\rm Strong} & 
$v_1 < c_{s1},\;\;v_2 > c_{s2}$ & 
$v_1 > c_{s1},\;\;v_2 < c_{s2}$ 
\\[2truemm] 
%
\end{tabular}
\vskip 0.8 truecm
\centerline{
\hskip 1.5truecm
\vbox{\hsize 12.0truecm\baselineskip=12pt\noindent\small
Table I.  The various combinations of the fluid velocities for the
different types of reaction front. The velocities are referred to the
front rest frame, with $v_1$ being the fluid velocity ahead of the front
and $v_2$ the fluid velocity behind the front. Similarly $c_{s1}$ and
$c_{s2}$ are the sound speeds on either side of the front.}}
\end{table}

	Because of their properties, reaction fronts have been considered
for studying the hydrodynamics of the phase interface during cosmological
first order phase transitions \cite{s82} such as the electroweak
\cite{eikr92} and the quark-hadron transition (see \cite{l94} for
detonations and \cite{bp93} for a review of deflagrations). In general,
first order cosmological phase transitions start with the nucleation of
bubbles of the low temperature phase within a supercooled ambient medium.
The surface that separates the two coexisting phases is then induced into
motion (the new phase is thermodynamically favoured) and in doing this it
transforms one phase into the other. An aspect which requires great
attention when studying the evolution of {\it thermodynamically stable}
reaction fronts is that of {\it hydrodynamic stability} and this
represents a large area of research both from the experimental and
the theoretical point of view. 

	The stability of classical deflagration fronts was first studied
by Landau in a seminal work of 1944 \cite{l44}. Following this, a number
of special relativistic linear stability analyses have been performed by
several authors both in the limit of small velocities \cite{l92} and in
the case of small and large velocities \cite{hkllm93}. These works had a
direct counterpart in the numerous studies produced on the hydrodynamics
of a cosmological first order phase transition in which the phase
interface moves as a weak deflagration front (see
\cite{gkkm84,mp90,ikkl94a,mr95} for the quark-hadron transition). 

	The results of these stability analyses have shown that, within a
linear theory, cosmological hadron bubbles could be unstable on time
scales much smaller than the typical time scale discussed for the duration
of the phase transition (the situation is different for the electroweak
transition, where the bubble wall seems stable under hydrodynamic
perturbations \cite{hkllm93}). It is well known however, that the ultimate
onset of the instabilities cannot be fully assessed within a linear
analysis, since it might also be that the instability modes are controlled
by intervening nonlinear effects which would limit the energy transfer
into the unstable modes. Of course, a necessary and sufficient condition
for the validity of the above arguments is that a mutual causal connection is
maintained between the front and the upstream or the downstream regions of
the flow. 
	
	The stability of classical detonation fronts has been studied
quite extensively in the one-dimensional and linear regime \cite{ls90} and
attempts are currently being made to extend this analysis within a weakly
nonlinear theory (see \cite{r91} for a list of references). A recent
investigation of the stability properties of relativistic detonation
fronts in cosmological phase transitions has been made by Abney \cite{a94}
who has concentrated particularly on the case of Chapman-Jouguet
detonations. As pointed out in \cite{a94}, while instability modes are not
allowed in the upstream flow region of a detonation front, it is not
possible to exclude them in the region behind the front, where they could
also grow exponentially with time. The present work aims to reconsider in
more detail the analysis performed in \cite{a94} both in the specific case
of Chapman-Jouguet detonations and in the general case of strong
detonations. 
	
	In order to do this, we start in Section II by presenting the
general equations deduced from the standard linear stability analysis of
special relativistic flows. In Section III we examine the stability of
generic discontinuity fronts with respect to corrugations and discuss the
application of these results to both Chapman-Jouguet and strong
detonations in Section IV. Section V is dedicated to the analysis of the
boundary conditions that detonation fronts need to satisfy in a
cosmological first order phase transition and there we comment on the
implications of the causal structure of weak detonations for their
stability properties. Conclusions are finally presented in Section VI. We
here adopt units for which $c = 1$, greek indices are taken to run from
$0$ to $3$ , while latin indices from $1$ to $3$. The metric has signature
$(-,\;+,\;+,\;+)$ and commas in covariant notation are used to denote
standard partial derivatives. 

{\who=1\bigskip\medskip\bigskip\goodbreak
{\Large\bf\noindent\hbox{II.}\hskip 0.5truecm {Linear hydrodynamic
stability}\nobreak \bigskip\medskip\nobreak\who=0}

	We here discuss the linear stability analysis of a relativistic
planar detonation wave. The procedure followed for the derivation of the
set of perturbed hydrodynamical equations parallels in part that presented
in \cite{a94} (where Chapman-Jouguet detonations only were considered),
but some important differences will emerge in the course of the
discussion. 

	Consider a plane detonation front which is propagating in a
Minkowski space-time with $(t,\, x,\, y,\, z)$ being the inertial
coordinates. The hydrodynamics of an unperturbed detonation wave can be
described as a one dimensional flow in which the whole space-time is
divided in two half spaces separated by a discontinuity surface moving at
four-velocity \hbox{$u^{\mu}_s=\gamma_s(1,\, v_s^i)$}, with
\hbox{$\gamma_s=(1-v^i_s \,v^j_s\, \delta_{ij})^{-1/2}$} and
\hbox{$v_s^i=dx_s^i/dt$}. 

	In this case, it is always possible to perform a Lorentz
transformation by means of which the front at any instant is taken to be
at rest on the $(y,\,z)$ plane and there are no three-velocity components
tangent to the front (hereafter we shall refer to the three-velocity
vectors simply as velocities). In this comoving reference frame, we can
think of the front as a planar discontinuity surface which divides the
three-space into the upstream region 1 and downstream region 2 and which
is crossed by positive velocities from left to right when the front is
left propagating in the inertial frame (see Figure 1). 

	Across  this surface matter undergoes either a combustion or a
phase transformation ({\it e.g.} passing from unconfined quarks to
light hadrons in the case of the quark-hadron phase transition).
Fluids on either side of the front are assumed to be ideal and
described by the standard stress-energy tensor of a relativistic
perfect gas 

\begin{equation}
\label{set}
T^{\mu \nu}=(e+p)u^{\mu}u^{\nu}+p g^{\mu \nu} \;,
\end{equation}

\medskip\noindent
 where $u^{\mu} \equiv \gamma(1\ ,\vec v)$ is the fluid four-velocity, $e$
is the energy density, $p$ is the pressure and $g^{\mu \nu}$ is the metric
tensor. The hydrodynamical equations can easily be derived from the
projection of the four-divergence of the stress-energy tensor along the
direction defined by the fluid four-velocity and orthogonal to it, so as
to express local conservation of energy and momentum respectively as

\begin{equation} 
\label{encons}
u_{\mu}T^{\mu \nu}_{\ \ \;,\>\nu}=0 \;,
\end{equation} 

\begin{equation} 
\label{momcons}
P_{\alpha \mu}T^{\mu \nu}_{\ \ \;,\>\nu}=0 \;,
\end{equation} 

\medskip
\noindent
where $P_{\mu \nu}=g_{\mu \nu}+u_{\mu} u_{\nu}$ is the projection operator
orthogonal to $u^{\mu}$ ({\it i.e.} \hbox{$P_{\mu \nu} u^{\nu}=0$}).
Making use of (\ref{set}), equations (\ref{encons}) and
(\ref{momcons}) can be written explicitly as 

\begin{equation} 
\label{encons1}
c^2_s w u^{\mu}_{\ ,\>\mu} + p_{,\>\mu} u^{\mu} =0 \;,
\end{equation} 

\noindent
and

\begin{equation} 
\label{momcons1}
w u_{\alpha,\>\mu} u^{\mu} + u_{\alpha}p_{,\>\mu} u^{\mu}+ 
p_{,\>\alpha} =0 \;,\end{equation} 

\noindent where $w=(e+p)$ is the enthalpy density and \hbox{$c_s=(\partial
p/\partial e)^{1/2}_s$} is the local sound speed (here $s$ is the specific
entropy). Equations (\ref{encons1}) and (\ref{momcons1}) represent the
usual ``zeroth order'' hydrodynamical equations and describe the motion of
fluid elements on either side of the detonation front. Following standard
linear perturbation analysis, we now introduce a small perturbation in the
relevant hydrodynamical variables so that, at first order in the
expansion, the new perturbed (primed) variables are

\begin{eqnarray}
\label {pert}
\hskip -1.0truecm p\ & \longrightarrow & p^{\prime} = p +\delta p \;,  
\\ \nonumber \\ \hskip -1.0truecm 
u^{\mu} \equiv \gamma(1,\;\vec v) = \gamma(1,\;v_x,\;0,\;0)\ &
\longrightarrow &
(u^{\mu})^{\prime} = u^{\mu} +\delta u^{\mu} \;, \hskip 2.0truecm 
\end{eqnarray}

\medskip
\noindent
where $\gamma = (1-v_x^2)^{-1/2}$ and (the flow is taken
to be uniform and unperturbed along the $z$-axis direction), 

\begin{equation}
(u^{\mu})^{\prime} \equiv \gamma(1+\gamma^2 v_x \delta v_x,\; 
v_x+\gamma^2 \delta v_x,\; \delta v_y,\; 0) \;. 
\end{equation}

\medskip
\noindent
 As a result, the perturbed expressions of equations (\ref{encons1}) and
(\ref{momcons1}) are

\begin{equation}
\label{hyd1}
c_s^2 w \left( \gamma^2 v_x {\partial \over{\partial t}} \delta v_x + 
\gamma^2 {\partial \over{\partial x}} \delta v_x + 
{\partial \over{\partial y}} \delta v_y \right) + 
{\partial \over{\partial t}} \delta p + 
v_x {\partial \over{\partial x}} \delta p =0 \;,
\end{equation}

\begin{equation}
\label{hyd2}
\gamma^2 w \left( {\partial \over{\partial t}} \delta v_x + 
v_x {\partial \over{\partial x}} \delta v_x 
\right) + v_x{\partial \over{\partial t}} \delta p + 
{\partial \over{\partial x}} \delta p = 0 \;,   
\end{equation}

\begin{equation}
\label{hyd3}
\gamma^2 w \left( {\partial \over{\partial t}} \delta v_y + 
v_x {\partial \over{\partial x}} \delta v_y 
\right) + {\partial \over{\partial y}} \delta p =0 \;,  
\end{equation}

\noindent
which can also be written, in a more compact form, as

\begin{equation}
\label{hydsyst}
\left( {\bf C}_t {\partial \over{\partial t}} + 
{\bf C}_x {\partial \over{\partial x}} + 
{\bf C}_y {\partial \over{\partial y}} \ \right) \vec{\bf U} = 0 \;,
\end{equation} 

\noindent
where 

\begin{equation}
{\bf C}_t = \left( 
\begin{array}{ccc}
1 & \gamma^2 w c_s^2 v_x & 0 \\ \\
v_x & \gamma^2 w & 0 \\ \\
0 & 0 & \gamma^2 w 
\end{array}
\right) \;,
\end{equation}

\medskip
\begin{equation}
{\bf C}_x = \left( 
\begin{array}{ccc}
v_x & \gamma^2 w c_s^2 & 0 \\ \\
1 & \gamma^2 w v_x & 0 \\ \\
0 & 0 & \gamma^2 w v_x
\end{array}
\right) \;,
\end{equation}

\medskip
\begin{equation}
{\bf C}_y = \left( 
\begin{array}{ccc}
0 & 0 & w c_s^2 \\ \\
0 & 0 & 0 \\ \\
1 & 0 & 0  
\end{array}
\right) \;,
\end{equation}

\noindent
and $\vec{\bf U}$ is the state-vector for the perturbations

\begin{equation}
\label{psv}
{\bf \vec U}=\left(
\begin{array}{c}
\delta p \\ \\
\delta v_x \\ \\
\delta v_y 
\end{array}
\right) \;.
\end{equation}
\medskip

	The most general solution of (\ref{hydsyst}) has the form

\begin{equation}
\label{sol}
\vec{\bf U}(t,\;x,\;y)=\vec{\bf A}(x)e^{-i(\omega t + k y)} \;, 
\end{equation}

\noindent
with $\omega$ being a complex number, $k$ a real number and

\begin{equation}
\label{ax}
\vec{\bf A}(x)=\sum_{j=1}^{3}{(c_j \vec{\bf L}_j) e^{-i(l_j x)} } \;.
\end{equation}

The $l_j$ are the complex eigenvalues of the secular equation
(\ref{hydsyst}), $\vec{\bf L}_j$ are the corresponding eigenvectors
and $c_j$ are three real constant coefficients. Substituting the trial
solution (\ref{sol}) in (\ref{hydsyst}), leads to an homogeneous system 
of equations whose coefficients are in the secular matrix 

\begin{equation}
{\bf D} \equiv ({\bf C}_t\ \omega + {\bf C}_x\  l + {\bf C}_y\ k)\;, 
\end{equation}

\noindent
 The eigenvalues $l_j$ can then be found by setting to zero the
determinant of ${\bf D}$. Doing this we obtain the dispersion relation

\begin{equation}
\label{eigenv}
{\rm det}\;({\bf D}) = (lv_x + \omega)\left[ (lv_x + \omega)^2 - 
(\omega v_x +l)^2 c^2_s  - (1-v_x^2) k^2 c^2_s  \right] =0 \;,
\end{equation}

\medskip\noindent
which has the roots

\begin{equation}
\label{l1}
l_1= - {\omega \over {v_x}} \;,
\end{equation}

\begin{equation}
\label{l23}
l_{2,3}= {1\over{v_x^2-c^2_s}} \left\{
(c^2_s-1)\omega v_x \pm c_s (1-v^2_x)\left[\omega^2+
{(v^2_x-c^2_s)\over{1-v^2_x}}k^2\right]^{1/2}\right\} \;.
\end{equation}

	A first point to notice is that the eigenvalues $l_{2,3}$ become
divergent for a fluid velocity normal to the front ($v_x$) equal to the
local sound speed. The presence of a singularity at the sonic point is a
general feature of the linear stability analysis of shock waves and reaction
fronts \cite{a89,ll93} and so it is necessary to take particular care when
examining these limiting cases. It is easy to realize however, that the
singularity in (\ref{l23}) is only an apparent one and it can be avoided
if one solves equation (\ref{eigenv}) directly with $v_x=c_s$. In this
case there are only two roots and the new eigenvalues (which we denote
with a bar) then are: $\bar l_1 = l_1$ and

\begin{equation}
\label{l2b}
\bar l_2={c_s k^2 \over {2 \omega}} -{(1+c^2_s)\over{2 c_s}} \omega \;.
\end{equation}

	Using the eigenvalue (\ref{l2b}) in the downstream region of the
flow solves the problem of the singularity at the sonic point, but
necessarily restricts the analysis to Chapman-Jouguet detonations only. In
order to investigate the stability properties of generic detonation
fronts, it is necessary to make use of the solutions (\ref{l1}) and
(\ref{l23}) for fluid velocities near to the sound speed. For this purpose
we can write the velocity normal to the detonation front as

\begin{equation}
\label{eps}
v_x = c_s+\epsilon \;,
\end{equation}

\medskip
\noindent 
 with $\epsilon$ being a small positive or negative real number. The value
of $\epsilon$ must in principle lie in the range \hbox{$0 < \epsilon <
(1-c_s)$} for the upstream region ($x<0$) and in the range \hbox{$-c_s <
\epsilon < (1-c_s)$} for the downstream region ($x>0$). Making use of
(\ref{eps}), it is possible to expand both the numerators and the
denominators of solutions (\ref{l1}) and (\ref{l23}) around the sonic
point so as to obtain the new eigenvalues (marked with a tilde)

\begin{equation}
\label{l1n}
{\tilde l}_1 \cong - {\omega \over {c_s+\epsilon}} \;,
\end{equation}

\begin{equation}
\label{l2n}
{\tilde l}_2 \cong - {1 \over {\epsilon(2c_s+\epsilon)}} 
\left\{ \omega \left(1+c^2_s-{k^2c^2_s\over{\omega^2}}\right)\epsilon +
\omega c_s\left[1-{k^2\over{2\omega^2}}+
{k^4 c^2_s \over{2\omega^4(1-c^2_s)}} 
\right] \epsilon^2 \right\} \;,
\end{equation}

\begin{eqnarray}
\label{l3n}
{\tilde l}_3 \cong {1 \over {\epsilon(2c_s+\epsilon)}} 
\Biggl\{ 2 \omega c_s(c^2_s-1)+ 
\omega \left(3c^2_s-1-{k^2c^2_s\over{\omega^2}} \right)\epsilon 
\hskip 3.0truecm  \nonumber \\
\hskip 5.0truecm  
+ \omega c_s\left[1-{k^2\over{2\omega^2}}+
{k^4 c^2_s \over{2\omega^4(1-c^2_s)}} 
\right] \epsilon^2
\Biggr\} \;.
\end{eqnarray}

	Although approximate, these expressions are suitable for a generic
value of the fluid velocity near the interface and provide a starting
point for the stability analysis of both strong and weak detonations. It
is important to notice that ${\tilde l}_2$ is not singular for $\epsilon
\rightarrow 0$ and that it reduces to $\bar l_2$ at first order; for this
reason a second order expansion is necessary. This is not the case for
${\tilde l}_3$ which represents the singular root of (\ref{l23}) and for
which the first order expansion is, in fact, sufficient. 

	Next, it is necessary to find the form of the eigenvectors
$\vec{\bf L}_j$ contained in (\ref{ax}). This requires solving the matrix 
equation

\begin{eqnarray}
\left( 
\begin{array}{ccc}
(\omega + {\tilde l}_j v_x) & \gamma^2 w 
(\omega v_x + {\tilde l}_j)c_s^2 & w c_s^2 k \\ \\
(\omega v_x + {\tilde l}_j) & \gamma^2 w (\omega + 
{\tilde l}_j v_x) & 0 \\ \\
k & 0 & \gamma^2 w (\omega + {\tilde l}_j v_x) 
\end{array}
\right) 
\left( 
\begin{array}{c}
\vec{\bf L}_{j\,1} \\ \\
\vec{\bf L}_{j\,2} \\ \\
\vec{\bf L}_{j\,3}  
\end{array}
\right) = 0 \;,
\nonumber \\
\nonumber \\ \hfill  j=1,\ 2,\ 3 \;,
\end{eqnarray}

\noindent
which leads to the following eigenvectors 

\begin{equation}
\label{bl1}
\vec{\bf L}_1 = \left( 
\begin{array}{c}
0\\ \\
1\\ \\
-\displaystyle {{\tilde l}_1\over k}
\end{array}
\right) \;,
\end{equation} 

\medskip
 \begin{equation}
\label{bl23}
\vec{\bf L}_{n} = \left( 
\begin{array}{c}
1\\ \\
\displaystyle{-{(\omega v_x + 
{\tilde l}_{n})\over{\gamma^2 w(\omega + {\tilde l}_{n}v_x)}}}\\ \\
\displaystyle{-{k \over{\gamma^2 w(\omega + {\tilde l}_{n}v_x)}}} 
 \end{array}
\right) \;, 
\end{equation} 
\vskip -0.5 truecm
\begin{eqnarray}
& \hskip 10.0 truecm  n=2,\;3  \nonumber 
\end{eqnarray} 
 
	It is now necessary to ascertain the values of the
coefficients $c_j$ for which the solution (\ref{sol}), with
eigenvalues \hbox{(\ref{l1n})--(\ref{l3n})} and eigenvectors
(\ref{bl1})--(\ref{bl23}), satisfies the necessary boundary
conditions. For this reason we have to check that, if there are time
growing instabilities, the effects of these should be limited in space
and not extend to infinity, so that 

\begin{equation}
\label{lim}
\lim_{x\to\pm\infty} |\vec {\bf U}(t,x,y)| = 0 \;.
\end{equation}

	After a few algebraic transformations, it is possible to verify
that

\begin{eqnarray}
\label{imw1}
{\rm (a)}\hskip 1.5truecm  
{\rm Im}\ (\omega) > 0& \ \ \Longleftrightarrow \ \ &{\rm Im}\ 
({\tilde l}_1) < 0 \;,
\nonumber \\
{\rm (b)}\hskip 1.5truecm  
{\rm Im}\ (\omega) > 0& \ \ \Longleftrightarrow \ \ &{\rm Im}\ 
({\tilde l}_2) < 0
\;, \nonumber \\
{\rm (c)}\hskip 1.5truecm  
{\rm Im}\ (\omega) > 0& \ \ \Longleftrightarrow \ \ &{\rm Im}\ 
({\tilde l}_3) < 0 
\hskip 1.0truecm  {\rm for}\  0 < \epsilon < (1-c_s)\ \;, 
\nonumber \\
\end{eqnarray}
\medskip
\noindent
and
\begin{eqnarray}
\label{imw2}
{\rm (d)}\hskip 1.5truecm  
{\rm Im}\  (\omega) > 0& \ \ \Longrightarrow \ \ &{\rm Im}\ 
({\tilde l}_1)< 0 \;, \nonumber \\
{\rm (e)}\hskip 1.5truecm  
{\rm Im}\  (\omega) > 0& \ \ \Longrightarrow \ \ &{\rm Im}\ 
({\tilde l}_2) < 0\;, \nonumber \\
{\rm (f)}\hskip 1.5truecm  
{\rm Im}\  (\omega) > 0& \ \ \Longrightarrow \ \ &{\rm Im}\ 
({\tilde l}_3)
> 0  \hskip 1.0truecm  {\rm for}\ -c_s < \epsilon \leq 0 \;, 
\nonumber \\
\end{eqnarray}

	In order for (\ref{lim}) to be satisfied, it is necessary that
\hbox{Im $({\tilde l}_j) > 0$} or that the corresponding coefficients
$c_j$ are zero for $x < 0$ and that \hbox{Im $({\tilde l}_j) < 0$} or
$c_j=0$ for $x > 0$. For modes with \hbox{Im $(\omega) > 0$} we then have

\begin{eqnarray}
\label{cj}
{\rm (a)}\hskip 0.75truecm
&c_1=c_2=c_3=0 \ \ &{\rm for}\ x < 0 \;, 
\nonumber \\ 
{\rm (b)}\hskip 0.75truecm
&c_1\not=0,\;\  c_2\not=0,\; \ c_3=0\ \ &{\rm for} \ x>0 \ \ 
{\rm and}\ \epsilon \leq 0 \; 
\hskip 1.0truecm {\rm (strong,\ C-J)\;,}
\nonumber \\
{\rm (c)}\hskip 0.75truecm
&c_1\not=0,\;\  c_2\not=0,\; \ c_3\not=0\ \ &{\rm for} \ x>0 \ \ 
{\rm and}\ \epsilon > 0 \; 
\hskip 1.0truecm {\rm (weak)}\;.
\nonumber \\
\end{eqnarray}

	In other words, the conditions (\ref{cj}) signify that no
perturbations can grow ahead of the detonation front ({\it i.e.} $\vec
{\bf U}(t, x, y) = 0$ for $x < 0$), while this is not necessarily the case
for the positive $x$ half-plane, where growing modes are allowed to exist
since only one coefficient needs to be zero in the case of strong and
Chapman-Jouguet detonations, and none in the case of weak detonations. 
This latter result represents an important difference between strong and
weak detonations and will be further underlined in the next Sections. The
condition (\ref{cj}-a) on the coefficients $c_j$ has its physical
interpretation in the fact that in the negative $x$ half-plane the flow is
supersonic and ``entering'' the front and, as a consequence, no sonic
signal (and therefore no perturbation) can be transmitted upstream of this
flow. In the next Section we shall study whether a dynamical evolution of
the instabilities behind the detonation front is possible when this is
subject to a corrugation instability. 

{\who=1\bigskip\medskip\bigskip\goodbreak
{\Large\bf\noindent\hbox{III.}\hskip 0.5truecm {Front corrugation 
stability}\nobreak \bigskip\medskip\nobreak\who=0}

	Within the theory of shock front stability, it is known that the
conditions for a shock to be evolutionary \hbox{({\it i.e.} $v_1>c_{s1}$,
$v_2<c_{s2}$)} \cite{ll93} are only necessary but not sufficient to prove
that it will not develop instabilities. This means that an evolutionary
shock ({\it i.e.} one for which any infinitesimal perturbation of the
initial state produces only an infinitesimal change in the flow over a
sufficiently short time interval), could become unstable (over a longer
time interval) with respect to small perturbations of the discontinuity
surface, which would then appear as ``corrugations'' of the front\footnote
{We note that it is possible, in principle, to write
		necessary and sufficient conditions for a 
		shock wave not to decay into a number of different 
		discontinuity surfaces \cite{gz87}. However, these 
		conditions do not provide information about the 
		evolution of the shock when this is subject to 
		corrugations of the front.} 
(see \hbox{Figure 1}). 

	Corrugation stability analyses of the type presented in this
paper have been performed in the past both for a non-relativistic
shock \cite{d54,ll93} and for a relativistic Chapman-Jouguet detonation
\cite{a94}. A rather different approach for a relativistic shock wave
has been proposed by Anile and Russo \cite{ar86, a89} where the
intuitive definition of corrugation stability introduced by Whitham
\cite{w74} has been translated into a more rigorous form (we recall
that according to Whitham, a corrugated shock front is stable if the
shock velocity decreases where the front in expanding and increases
where it is contracting). 

	We here discuss the corrugation stability of strong relativistic
detonation fronts and show how these relate to the special case of
Chapman-Jouguet detonations. The first step consists in establishing the
correct eigenvalues to choose. Making use of the conditions
(\ref{imw1})--(\ref{imw2}), together with (\ref{cj}), it is evident that
it is necessary to use the eigenvalues ${\tilde l}_1,\; {\tilde l}_2$ (and
the corresponding eigenvectors) in the case of strong and Chapman-Jouguet
detonations ($\epsilon \leq 0$) while all of the eigenvalues ${\tilde
l}_1,\;  {\tilde l}_2,\;{\tilde l}_3$ need to be used in the case of weak
detonations ($\epsilon>0$). The second step consists of requiring that the
perturbed hydrodynamical equations satisfy junction conditions at the
phase interface expressing the conservation of energy and momentum
respectively. 

\noindent
In a Minkowski spacetime these reduce to (see \cite{mp89,rm94,rm95}
for a general relativistic treatment) 

\begin{equation}
\label{conse}
\left[ \gamma^2 w v_x \right]_{1,2}=0 \;,
\end{equation}

\begin{equation}
\label{consm}
\left[\gamma^2 w v^2_x + p \right]_{1,2}=0 \;,
\end{equation}

\begin{equation}
\label{constv}
\left[ v_y \right]_{1,2}=0 \;,
\end{equation}

\noindent
where $[A]_{1,2}=A_1-A_2$ (note that these junction conditions need
always to be expressed in the front rest frame and that the latter
coincides with the coordinate frame only when the front is
unperturbed). We next perturb the front with a periodic oscillation in
the \hbox{$y$-axis} direction of the type (see Figure 1)

\begin{equation}
\label{delta}
\Delta=\Delta_0 e^{-i(\omega t + k y)} \;,
\end{equation}

\noindent 
and calculate the resulting form of the perturbed junction conditions.
For this purpose it is convenient to introduce orthogonal unit
three-vectors normal ($\underline {\bf n}$) and tangent ($\underline
{\bf t}$) to the front (Figure 1) which, at the first order in the
perturbation, have components 

\begin{equation}
\underline{\bf n}\equiv (1,\;-{\partial \Delta \over{\partial y}},\;0)=
(1,\;ik\Delta,\;0)\;, \hskip 1.5 truecm 
\underline{\bf t}\equiv ({\partial \Delta \over{\partial y}},\;1,\;0)=
(-ik\Delta,\;1,\;0)\;.  
\end{equation}

\bigskip
\bigskip
\centerline{{\psfig{figure=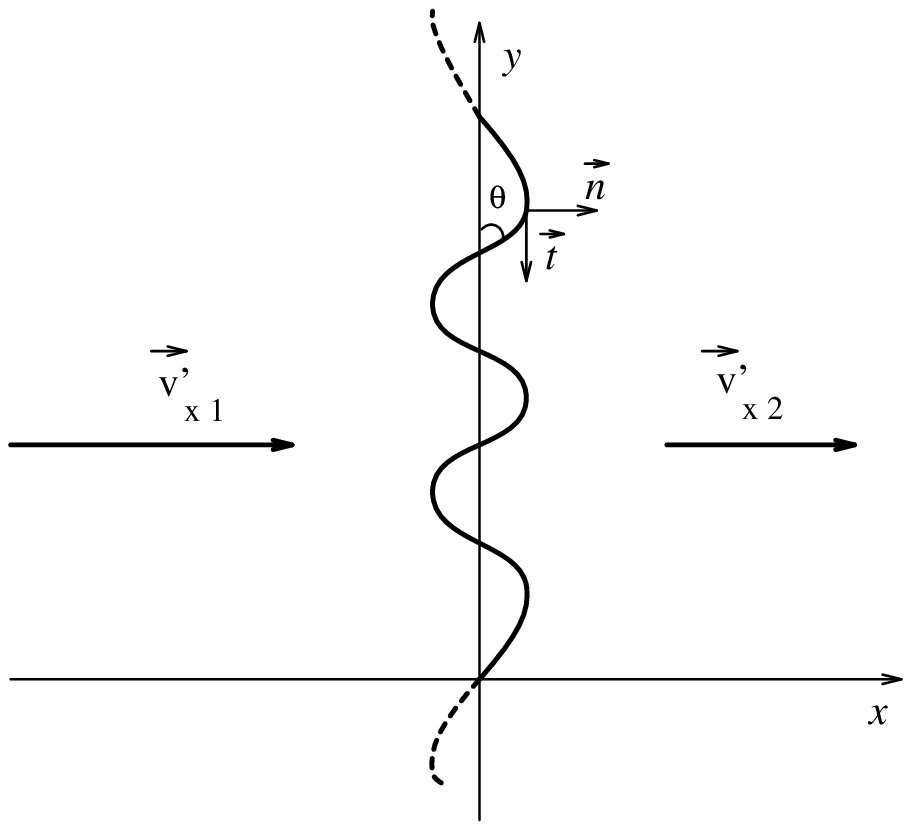,height=8.0truecm,width=10.0truecm}}}
\medskip
\vskip 0.5truecm
\centerline{\vbox{\hsize 12.0truecm\baselineskip=12pt\noindent\small 
Figure 1.  Schematic representation of a corrugated detonation front.
\hbox{$\underline {\bf n}$ and $\underline {\bf t}$} are orthogonal
unit three-vectors, normal and tangent respectively to the
discontinuity surface. }} 
\bigskip
\bigskip

	It is now necessary to evaluate the perturbed expressions for
the fluid velocities on either side of the front, as these are viewed
in the front rest frame. For this purpose it is necessary to perform a
relativistic velocity transformation with respect to the detonation
velocity \hbox{${\vec v}_f=(\partial \Delta /\partial t)\>
\underline{\bf n}=(-i \omega \Delta) \> \underline{\bf n}$}, so as to
obtain the following expressions for the perturbed normal and
tangential velocities relative to the front 

\begin{equation}
\label{vnx}
 {\vec v}^{\; \prime}_j \cdot \ \underline{\bf n} =
\left( {{v_{xj} + \delta v_{xj} + i \omega \Delta }\over 
{1 + i \omega v_{xj} \Delta }}
,\> \delta v_{yj},\> 0 \right) \cdot \ \underline{\bf n} \cong
v_{xj} + \delta v_{xj} + 
\displaystyle {i \omega \Delta \over {\gamma^2_j}}\;, 
\end{equation}

\begin{eqnarray}
\label{vny}
& {\vec v}^{\; \prime}_j \cdot \ \underline{\bf t} \cong
\delta v_{yj} - i k v_{xj} \Delta \;, \\ 
& \hskip 8.0 truecm j=1,\> 2 \;\;, \nonumber 
\end{eqnarray}

\noindent
and to which corresponds a perturbed gamma factor

\begin{eqnarray}
\label{ngf}
&\delta \gamma_j \cong \gamma^3_j v_{xj} \left( \delta v_{xj} + 
\displaystyle {i \omega \Delta \over {\gamma^2_j}} \right) \;, \\
&\hskip 8.0 truecm j=1,\> 2 \;\;. \nonumber 
\end{eqnarray}

	It is now convenient to introduce the new state-vector of the
perturbations near to the interface and in downstream region of the flow 

\begin{equation}
\label{npsv}
{\bf \vec V}_2(t,\;y)=\left(
\begin{array}{c}
\delta p_2 \\ \\
\delta v_{x2} \\ \\
\delta v_{y2} 
\end{array}
\right) \;,
\end{equation}

\medskip 
\noindent 
where, as discussed in the previous Section, each component of the
corresponding ${\bf \vec V}_1(t,\>y)$ is automatically zero. Making use of
(\ref{vnx})--(\ref{ngf}) in (\ref{conse})--(\ref{constv}), it is possible
to write out the perturbed junction conditions and to use the resulting
system of three equations in order to derive the components of ${\bf \vec
V}_2(t,\>y)$. In doing this, we should also modify the momentum balance
across the interface by taking into account the response of the front
which is due to its surface tension $\sigma$ (the energy balance at the
front is unaffected by a constant surface tension \cite{m86,mp89}). This
contribution appears in the expressions for the negative $x$ half-plane
with the form \hbox{$\sigma(\partial^2/\partial y^2 - \partial^2/\partial
t^2) \Delta$}, where the first term is related to the surface curvature,
while the second is related to its ``inertia''. Omitting here the lengthy
algebra, the three components of ${\bf \vec V}_2(t,\;y)$ are found to be

\begin{equation}
\label{dp2}
\delta p_2 = {1+v^2_2\over{\Gamma_-+\Gamma_+ v^2_2}}
\left[ 2 i \omega w_2 {(v_1-v_2)(1-v_1v_2)\over{(1+v^2_2)}} 
{\gamma^2_2 v_2\over{v_1}} + \sigma (\omega^2-k^2) \right] \Delta \;,
\end{equation}

\begin{equation}
\label{dvx2}
\delta v_{x2} = - {1-v^2_2\over{\Gamma_-+\Gamma_+ v^2_2}}
\left[ i \omega {(v_1-v_2)(1-v_1v_2)\over{v_1}} \Gamma_+ + 
\sigma (\omega^2-k^2) {\Theta_2 \over{w_2}} v_2 \right] \Delta \;,
\end{equation}

\begin{equation}
\label{dvy2}
\delta v_{y2} = - i k \Delta (v_1-v_2) \;,
\end{equation}

\medskip 
\noindent 
where \hbox{$\Gamma_{\pm}= (1 \pm \Theta_2 \gamma^2_2 v^2_2)$}, with
\hbox{$\Theta_2=(\delta w_2/\delta p_2)=1+1/c^2_{s_{2}}$} for a
relativistic gas (note that for compactness we will write $v_{1,2} \equiv
v_{x1,x2}$ for the zeroth order velocities). Expressions
(\ref{dp2})--(\ref{dvy2}) represent the special relativistic
generalization of the equivalent expressions discussed in \S 90 of
\cite{ll93} and reduce to these when the Newtonian limit is taken. It is
important to remark that the term \hbox{$(\Gamma_-+\Gamma_+
v^2_2)=1+v^2_2(1-\Theta_2)$}, in the denominators of
(\ref{dp2})--(\ref{dvx2}), vanishes whenever \hbox{$v_2=c_{s2}$}, giving a
singular behaviour at the sonic point similar to the one seen in the
Newtonian case \cite{ll93}. We note that the corresponding expressions
derived in \cite{a94}, differ from (\ref{dp2})--(\ref{dvx2}) and do not
show this singular behaviour except in their Newtonian limit. This
discrepancy is due to the fact that in Abney's treatment the
transformation to the front rest frame is a simple Galileian one and that,
also, the perturbation in the squared gamma factor is neglected.
Unfortunately, we believe that these omissions, which radically change
the nature of the solution at the sonic point and strongly influence the 
analysis, cannot be considered satisfactory. 

	As mentioned in Section II, a singular behaviour for a velocity
behind the front equal to the local sound speed is a standard feature of
the stability analysis of discontinuity surfaces. Nevertheless, great care
must be taken when discussing these limiting cases, such as the present
Chapman-Jouguet detonations. Some physical insight into the properties of
Chapman-Jouguet detonations can already be gained when looking at the
perturbed expressions for the hydrodynamical variables in terms of the
perturbation $\Delta$ [{\it i.e.} inverting expressions
(\ref{dp2})--(\ref{dvx2})]. 

	In this case, in fact, we could conclude that the corrugations
produced on a Chapman-Jouguet detonation front by perturbations in the
downstream fluid variables, are always zero and independent of the
strength of the perturbations ({\it i.e.} independent of the magnitude of
${\bf \vec V}_2$). In other words, expressions (\ref{dp2})--(\ref{dvx2})
seem to indicate that at linear order a Chapman-Jouguet detonation is {\it
unconditionally stable} (an identical conclusion will be drawn also from
the study of the dispersion relation for a Chapman-Jouguet detonation in
Section IV). 

	At this stage it is possible to deduce the form of the dispersion
relation by requiring that the hydrodynamical perturbations present in the
fluid adjacent to the phase interface are compatible and coincide with the
perturbations produced by the corrugations of the front, {\it i.e.}

\begin{equation}
\label{u=v}
{\bf \vec U}(t,\;0^+,\;y) \equiv
\sum_{j=1}^3 (c_j \vec {\bf L}_j)e^{-i(\omega t + k y)} =
{\bf \vec V}_2(t,\;y) \;. 
\end{equation}

\noindent
 Writing out (\ref{u=v}) explicitly results in a system of three equations
with unknowns being given by the coefficients $c_j$, and by the surface
displacement $\Delta_0$. Whether equations (\ref{u=v}) are sufficient to
determine the dispersion relation depends on the number of nonzero
coefficients $c_j$ or, equivalently, on the degree of ``underdeterminacy''
of a detonation front. This concept appears in a simple procedure which is
used for discussing the stability properties of reaction fronts or
discontinuity surfaces in general \cite{ll93,f78}. The degree of
``underdeterminacy'' of a discontinuity can be calculated by counting the
number of free parameters which could be associated with a small
perturbation of the front (these are given by the number of sonic
perturbations transmitted from the front\footnote{
		In Figure 3 the number of sonic perturbations can be
		counted by using the characteristic curves for the 
		different fronts summarized in Table I. }, the entropy  
perturbation propagated in the downstream region of the flow and the
surface displacement) minus the number of boundary conditions that the
perturbation has to satisfy (in general there are three of these:
conservation of mass, energy and momentum). 

	In the case of a strong or Chapman-Jouguet detonation there exist
three free parameters (which correspond to the unknowns $c_1,\; c_2$ and
to $\Delta_0$) and the front is then with zero degree of
``underdeterminacy''. In this case, equation (\ref{u=v}) has a solution
provided that the determinant of the matrix of coefficients vanishes, {\it
i.e.}

\begin{equation}
\label{det=0}
{\rm det}\ \left( 
\begin{array}{ccc}
0 & 1 &\delta p_2 \\ \\
1 & \displaystyle{-{(\omega v_x + {\tilde l}_{2})
\over{\gamma^2_2 w_2(\omega + {\tilde l}_{2}v_x)}}}
&\delta v_{x2} \\ \\
-\displaystyle {{\tilde l}_1\over k} & 
\displaystyle{-{k \over{\gamma^2_2 w_2(\omega + 
{\tilde l}_{2}v_x)}}} & \delta v_{y2}
\end{array}
\right) = 0\;, \nonumber \\
\end{equation} 

\medskip\noindent
 After some algebra, the general form of the dispersion 
relation is found to be 

\begin{eqnarray}
\label{dr}
{w_2\over{\Gamma_-+\Gamma_+v^2_2}}\Biggl\{i{(v_1-v_2)\over{v_1}}
\Biggl[\omega^3 {(1-v_1v_2)(\Gamma_+-2\gamma^2_2 v^2_2)\over{v_2}}+
\omega^2 (1-v_1v_2)(\Gamma_+ - 2\gamma^2_2){\tilde l}_{2} 
\nonumber \\
+\omega \left[2v_2(1-v_1v_2)-
v_1 \left(\Gamma_-+\Gamma_+v^2_2\right)\right]\gamma^2_2 k^2 -
\gamma^2_2 v_1v_2 k^2 \left(\Gamma_-+\Gamma_+v^2_2\right) 
{\tilde l}_{2}\Biggr]  \hskip 1.0truecm \nonumber \\ 
+\sigma {(\omega^2 -k^2)\over{w_2}}\left[ \omega^2
\left(\Theta_2-1-v^2_2\right)
-{\omega \over{v_2}}\left(\Gamma_-+\Gamma_+v^2_2\right) 
{\tilde l}_{2} +
(1+v^2_2) k^2 \right]\Biggr\} = 0 \;, 
\nonumber  \\
\end{eqnarray}

\noindent 
which provides a relation $\omega=\omega(k)$ once the free variables
$v_1,\> v_2,\> c_{s2}$ and $\sigma/w_2$ are specified. 

	On the other hand, in the case of a weak detonation there exist
four free parameters (corresponding to the unknowns $c_1,\; c_2,\; c_3$
and $\Delta_0$) and this forces the introduction of a fourth boundary
condition in order to make the solution evolutionary. In this respect,
weak detonations are similar to weak deflagrations and in order to be
fully determined require an equation describing the microscopic burning
mechanism or, in the case of phase transitions, the rate of transformation
of the old phase into the new one. This feature of weak detonations does
not allow for a general discussion of their stability properties and
restricts the analysis to the specific situations in which the fourth
boundary condition can be expressed. For this reason, we will limit
ourselves to discuss the general stability properties of strong and
Chapman-Jouguet detonations only, remainding the study of weak detonations
in cosmological phase transitions to a future work \cite{r2000}. 

The next Sections will present the solution of (\ref{dr}) for the
different cases of Chapman-Jouguet and strong detonations. 

{\who=1\bigskip\medskip\bigskip\goodbreak
{\Large\bf\noindent\hbox{IV.}\hskip 0.5truecm {Chapman-Jouguet and 
strong detonations}\nobreak \bigskip\medskip\nobreak\who=0}

	We here discuss the problem of potential corrugation instabilities
behind the front of Chapman-Jouguet and strong detonations.  Recalling the
definition (\ref{l2n}) of ${\tilde l}_2$, it is possible to see that a
strong detonation naturally evolves into a Chapman-Jouguet detonation when
the velocity behind the front passes from being subsonic to being equal to
the sound speed. 

	We shall start by discussing this latter case, for which
$\epsilon=0$, $v_2=c_s$ (hereafter \hbox{$c_s\equiv c_{s2}$}) and
(\ref{dr}) reduces to

\begin{eqnarray}
\label{drcj}
{w_2\over{\Gamma_-+\Gamma_+c^2_{s}}}\Biggl\{i{(v_1-c_{s})\over{v_1}}
\left[ 2 \omega^3 {(1-v_1c_{s})\over{c_s}}+
2 \omega (1-v_1c_{s})\gamma^2_2 c_{s} k^2\right] \nonumber \\
+\sigma {(\omega^2 -k^2)\over{w_2}}
\left[ \omega^2 {(1+c^2_{s})\over{\gamma^2_2 c^2_{s}} } +
(1+c^2_{s})k^2 \right]\Biggr\} = 0 \;.
\end{eqnarray}

	Note that the dispersion relation (\ref{drcj}) does not 
contain the eigenvalue ${\tilde l}_2$ (which is always multiplied by 
vanishing terms) and that, in order to avoid a singularity, the
content of the curly brackets in (\ref{drcj}) has to be zero. 
This condition can be imposed by requiring that 

\begin{equation}
\label{drcjf}
(\omega^2 + \gamma^2_2c^2_{s}k^2)\left[2 i \omega 
{(v_1-c_s)(1-v_1c_s)\over{v_1c_s}}+
\sigma {(\omega^2 -k^2)\over{\gamma^2_2 c^2_s w_2}}\right]=0 \;,
\end{equation}

\medskip
\noindent
which has the four distinct roots 

\begin{equation}
\label{cjw12}
\omega_{1,2}=\pm \ i\gamma_2 c_s k  \;, 
\end{equation}

\begin{eqnarray}
\label{cjw34}
\omega_{3,4}=-{\gamma^2_2 c_s w_2 \over{v_1\sigma}}
\Biggl\{i(v_1-c_s)(1-v_1c_s) \mp \hskip 4.0truecm \nonumber \\
\left[\left( {v_1 \sigma \over{\gamma^2_2 c_s w_2}}\right)^2 k^2
-(v_1-c_s)^2(1-v_1c_s)^2\right]^{1/2}\Biggr\} \;.
\hskip -2.0 truecm \nonumber \\
\end{eqnarray}
	
	We note that it is a common experience in dealing with the
solution of dispersion relations that spurious roots can be introduced,
which then need to be discarded on the basis of physical or conceptual
considerations. An example of this is the root $\omega_1$ which has
positive imaginary part and would lead to an exponentially growing
unstable mode. However, $\omega_1$ should be rejected since it does not
satisfy energy boundary conditions at short wavelengths and would
lead to a ``high frequency catastrophe''. It is well known, in fact, that
the surface energy associated with a perturbation of amplitude $\Delta$ is
proportional to $\sigma k^2 \Delta^2$ and a cut-off wave number, above
which instabilities are not allowed, is necessary in order to avoid
accumulation of infinite energies at high frequencies. 
	
	The physical mechanisms which operate this limitation depend on
the specific situation under examination and can be either dissipative
effects, such as a fluid viscosity, or can be the consequence of surface
tension. However, the root $\omega_1$ does not contain any contribution
coming from the surface tension and this has the consequence that even an
infinitely stiff front ({\it i.e.} one with \hbox{$\sigma \rightarrow
\infty$}) would appear to be unstable at all wavelengths. This behaviour
suggests that the roots $\omega_{1,2}$ cannot provide a physical
description of detonation fronts and will be discarded here. Further
evidence of the inapplicability of the roots (\ref{cjw12}) comes from
realizing that the term \hbox{$(\omega^2+\gamma^2_2c^2_{s}k^2)$} in
(\ref{drcjf}) is the consequence of a Doppler frequency transformation
relative to a medium moving at the sound speed. With a few simple
calculations it is easy to see that this term should always be different
from zero (see the Appendix). 

	After some algebraic manipulations (the details of which can be
found in the Appendix), it is easy to show that the other two roots
$\omega_{3,4}$, which are clearly dependent on $\sigma$, both have
negative imaginary parts and therefore lead to {\it stable} solutions. As
discussed in Section III, this is further and more direct proof that at
first order a Chapman-Jouguet detonation is {\it unconditionally stable}. 
We remark that this conclusion is in contrast with the results presented
by Abney in \cite{a94} which indicated that Chapman-Jouguet detonations
are effectively unstable at all wavelengths. As mentioned in the previous
Section, the origin of this discrepancy is to be found in the
approximations adopted in \cite{a94} for the derivation of the perturbed
hydrodynamical quantities behind the detonation front, which have
artificially removed the singular properties of this class of detonations. 

	The situation is not very different when the more general
strong detonations are discussed. In this case, all of the expressions in
the dispersion relation (\ref{dr}) need to be expanded around the
sound speed, with terms up to the second order being retained. This is
a consequence of the fact that at first order Chapman-Jouguet
detonations and strong detonations are indistinguishable and a second
order expansion is therefore necessary. The complete general
dispersion relation, which results from lengthy algebraic manipulations,
is presented in the Appendix. Here, we limit ourselves to discussing 
the equivalent expression obtained by setting $c^2_s=1/3$ 

\begin{eqnarray}
\label{drsd}
\Bigg\{
{2 \sigma \over{w_2}}\left({4\over 3} -7 c_s \epsilon +
4 \epsilon^2 \right) \omega^7 
+i \left[8c_s- v_1\left(1+{1\over{v^2_1}}\right)(2+3c_s \epsilon)
+12 c_s \epsilon^2\right]\omega^6 
 \nonumber \\
-{\sigma k^2 \over{w_2}}\left({4\over 3} -19c_s \epsilon + 
{17\over 2}\epsilon^2 \right)\omega^5
+ik^2\Biggl[4c_s-v_1\left(1+{1\over{v^2_1}}\right)
 \hskip 2.0 truecm \nonumber \\
+\left(10-{3 v_1 c_s\over 2}
\left(5+{7\over{v^2_1}}\right)\right)\epsilon 
+\left(54c_s -{69\over 4}
\left(1+{1\over{v^2_1}}\right)\right)\epsilon^2 
\Biggr] \omega^4
 \hskip 1.0 truecm \nonumber \\
-{\sigma k^4\over{w_2}} \left({4\over 3}+5c_s\epsilon + 
{\epsilon^2\over 4}\right)\omega^3
-i{k^4 \epsilon \over 2}\left[1-3 v_1 c_s -
\left({v_1\over 4}\left(11+{1\over{v^2_1}}\right)-
{13\over 2}c_s\right)\epsilon \right]\omega^2
\nonumber \\
+{3 k^6 \sigma \epsilon^2 \over {4 w_2}}\omega 
+ i{3 k^6 \over 8}(c_s-v_1)\epsilon^2 \Biggr\}
{w_2\over{3 \epsilon (2c_s+\epsilon)}}= 0
\;. \hskip 2.0truecm \nonumber \\  
\end{eqnarray}

	As for the case of (\ref{drcj}), equation (\ref{drsd}) can be
satisfied only if the content of the curly brackets is set equal to zero and
this results in a seventh order polynomial in $\omega$ with complex
coefficients. (It is straightforward to check that (\ref{drsd}) reduces to
(\ref{drcjf}) when $\epsilon=0$.) The roots of (\ref{drsd}) can only be
computed numerically and for this we have implemented an algorithm which
makes use of a variant of Laguerre's Method (NAG Fortran Library C02AFF
\cite{nag91}). The computations can be performed only after all the
parameters have been specified and for this purpose we have set
$v_1=(3/2)c_s$ and \hbox{$\epsilon=-0.01$}. We stress that only the
solution of the complete polynomial allows one to deduce a consistent
picture of the functional dependence of the growth rate on the wavenumber
$k$. Any analysis which studies the dispersion relation (\ref{drsd})
in the approximate expressions which it assumes in the long and short
wavelength {\it limits} \cite{a94} can easily give rise to confusing
outcomes. 

	In analogy with \cite{a94}, we present in Figure 2 (a) results of
the numerical computations for the long wavelength region and in Figure 2
(b) for the short wavelength one, with the wavenumber being expressed in
units of $w_2/\sigma$.  Figures 2 indicate that there is always a root of
(\ref{drsd}) which has positive imaginary part (all the others have either
negative or constant imaginary part). 

	The instabilities described by this root have a growth rate which
is independent of $\sigma$, increases linearly with the wavenumber $k$ and
produces an inevitable energy divergence at high frequencies. They clearly
represent the relics of the unstable modes contained in the root
$\omega_1$ of the dispersion relation (\ref{drcjf}) and in analogy with
what deduced for Chapman-Jouguet detonations, the results shown in Figures
2 refer to a physically inconsistent solution and should be discarded. 

	As a result, we are led to conclude that strong detonations are
linearly {\it stable} to corrugations of the front and in this they
resemble Chapman-Jouguet detonations to which they reduce for
\hbox{$\epsilon \rightarrow 0$}. A similar result ({\it i.e.} that an
evolutionary front is also stable with respect to corrugations), has been
obtained also by Anile and Russo \cite{ar87,a89,ar86} in the context of
the stability of shock waves. Proceeding within the theory of singular
hypersurfaces, they came to the conclusion that the linear stability
conditions for planar relativistic shock waves coincide with those
obtained in the framework of corrugation stability. 

\vfill\eject

\null\vskip -2.5truecm
\hskip 1.5truecm \hbox{(a)}

\centerline{{\psfig{figure=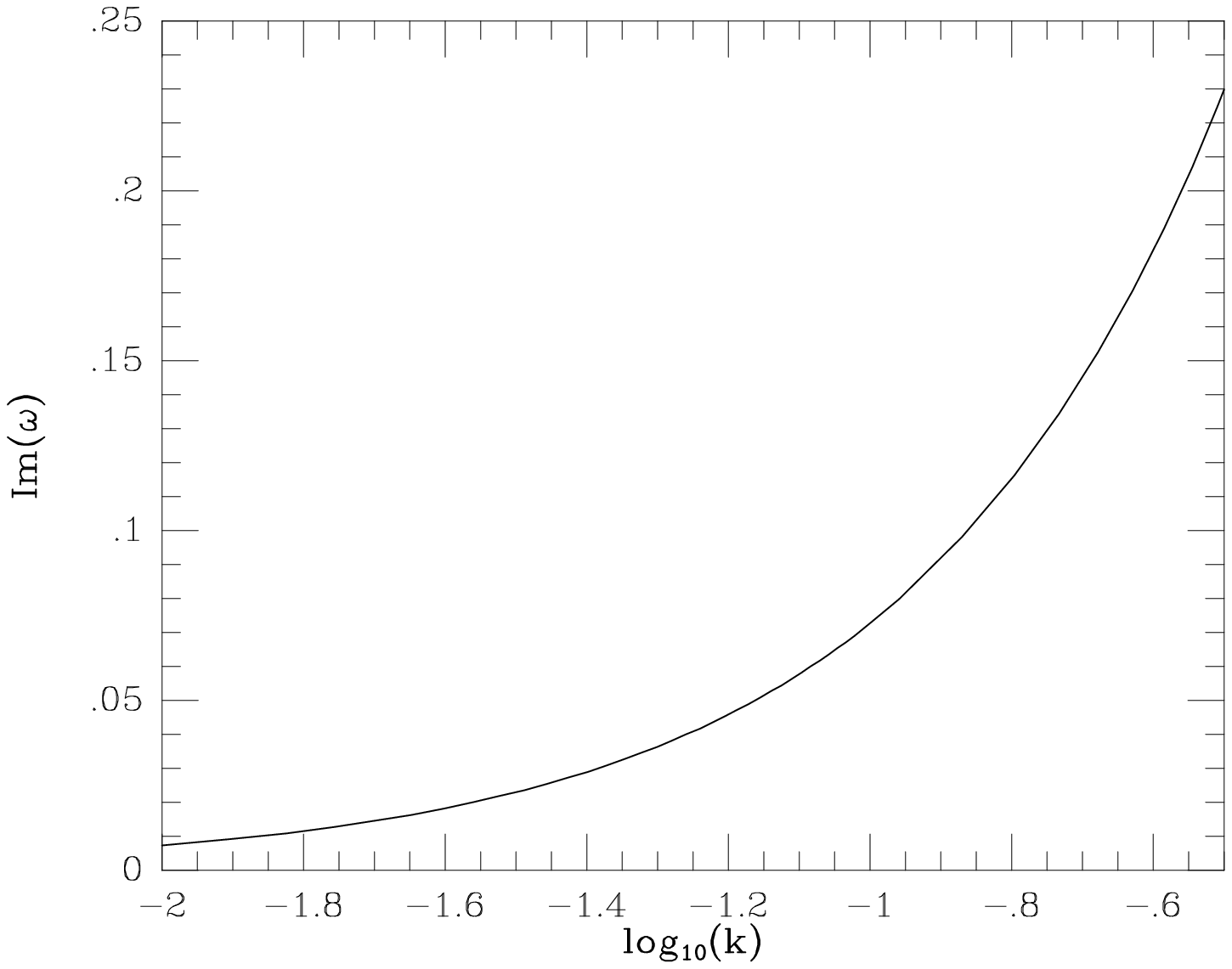,height=8.0truecm,width=10.0truecm}}}
\vskip 0.5truecm
\hskip 1.5truecm \hbox{(b)}

\centerline{{\psfig{figure=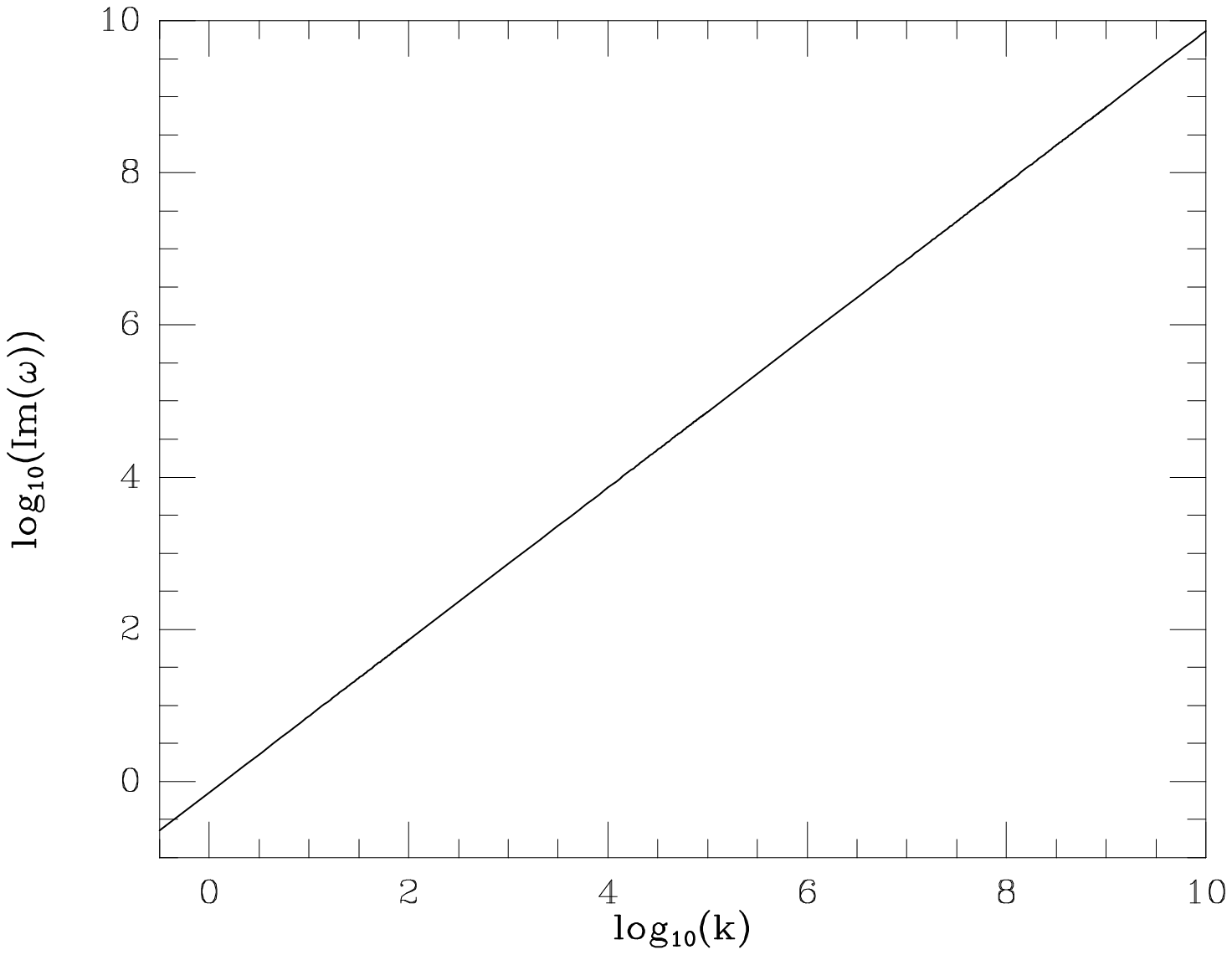,height=8.0truecm,width=10.0truecm}}}
\vskip 0.45truecm
\centerline{\vbox{\hsize 13.0truecm\baselineskip=10pt\noindent\small
Figures 2.  Perturbation growth rate Im $(\omega)$ plotted as a function
of the logarithm of the real wavenumber $k$ expressed in units of
$w_2/\sigma$. Figure 2 (a) refers to the long wavelength region, while
Figure 2 (b) to the short wavelength one. The curves have been calculated
assuming \hbox{$v_1=$ (3/2) $c_s$}, $c_s=1/\sqrt 3$ and \hbox{$\epsilon=-$ 
0.01}. }} \bigskip 
\bigskip
\vfill\eject

	In the next Section we shall show that, in spite of these
stability results, strong detonations cannot satisfy elementary boundary
conditions and for this reason they should be ruled out from the scenario
of cosmological first order phase transitions. 

{\who=1\bigskip\medskip\bigskip\goodbreak
{\Large\bf\noindent\hbox{VI.}\hskip 0.5truecm {Detonation fronts in 
cosmology }\nobreak \bigskip\medskip\nobreak\who=0}

	We here briefly investigate the consequences of the stability
analysis performed in the previous Sections for a cosmological first order
transition (as the electroweak or the quark-hadron) taking place by means
of bubbles of the new phase growing as detonations moving into the old
phase medium. It might be relevant to underline that the results of the
stability analysis can find a consistent application only when the bubbles
are sufficiently large so that they can be considered locally planar and
this is the assumption we will follow hereafter. 

	Firstly, we consider the situation of a spherical bubble moving as
a strong detonation; the results presented in Section IV indicate that
these reaction surfaces are both evolutionary and stable with respect to
corrugation instabilities. However, it is possible to demonstrate that
such flow configurations cannot be realized during bubble growth as they
cannot satisfy the required boundary conditions. Proofs of this have been
given in \cite{ll93} for nonrelativistic planar and spherical fronts, in
\cite{s82} for relativistic spherical fronts and also in \cite{l94} for
the case of relativistic planar fronts. The impossibility of having this
class of reaction front can be shown also with more simple arguments.
Consider a fluid element which is immediately behind a strong detonation
and which has been just transformed into the new phase (we here make the
implicit assumption that the sound speeds are constant on either side of
the front). This fluid element, which was initially at rest, has been put
into motion by the front which will be seen by the fluid element as
receding at a subsonic velocity. Symmetry and the presence of an origin
for the bubble, require that the fluid velocity behind the front becomes
zero somewhere in the flow profile and this could occur either via a
rarefaction wave, or via a shock wave. However, the front edge of the
rarefaction wave or the discontinuity surface would be seen as moving at
the sound speed or faster relative the fluid element and as a consequence
either one would inevitably overtake the detonation front. As a result,
neither of these two flows can be established behind a strong detonation
front which is then unable to adjust itself to the required boundary
conditions and so cannot be produced in practice. 

	As mentioned in Section IV, Chapman-Jouguet detonations represent
a limiting case of the more general strong detonations and share with the
latter the properties of stability with respect to corrugations of the
front. In the case of Chapman-Jouguet detonations however, the boundary
conditions for the medium behind can be suitably satisfied (the front is
moving at the sound speed as seen from the medium behind) and it is easy
to show that the detonation front is then followed by a rarefaction wave
which progressively decelerates and decompresses the fluid in the new
phase \cite{ll93,cf76,s82}. The occurrence of cosmological Chapman-Jouguet
detonations is however made difficult because of the considerable amount
of supercooling required before the nucleation of the new phase bubbles
and because they would tend to leave the new phase in a superheated state
\cite{bp93,gkkm84,ikkl94a}. 

	Next, we consider the case of a phase interface in a cosmological
phase transition moving as a weak detonation. This scenario has recently
been re-examined by Laine \cite{l94,kl95} who has proposed the possibility
of considering such fronts as a ``natural mechanism for bubble growth in
phase transitions'' in view of the differences between chemical burning
(in which context they are excluded) and cosmological phase transitions. 

	Boundary conditions for a weak detonation can be easily satisfied
since the flow behind the front can be suitably ``adjusted'' by means of a
rarefaction wave which progressively slows down the new phase, bringing it
to rest at the rear edge of the wave which behaves as a weak discontinuity
moving at the local sound speed. Moreover, weak detonations require a
degree of supercooling smaller than the one needed by Chapman-Jouguet
detonations and would also produce a smaller or zero superheating of the
low temperature phase. As discussed in Sections II and III, a general
discussion of the stability properties of weak detonations is not possible
because of the intrinsic ``underdeterminacy'' of these fronts. However, we
want to recall the attention on a general a feature of the causal
structure of weak detonations that should be taken into account when
performing a stability analysis. 

	Generally speaking, it is not implausible to expect that
higher order effects could intervene during the growth of linear
instabilities so as to modify the energy balance and saturate the
oscillations which could then be present but would not grow unboundedly
with time. A similar argument is rather appropriate for many instability
phenomena and has been used also in relation to cosmological deflagrations
\cite{l92} and Chapman-Jouguet detonations \cite{a94}. It is essential to
stress however, that a non-linear saturation produced by mutual
interaction between the front and the adjacent fluid cannot be invoked in
the case of weak detonation fronts, because of the intrinsic causal
structure of a weak detonation. 

	In order to clarify the arguments, in Figure 3 we have shown
schematically the causal structure of the six types of reaction front
which were summarized in Table I. The figure consists of six different
spacetime diagrams, (all referred to the front rest frame), with time on
the vertical axis and the space coordinate on the horizontal one. For each
diagram, the thick solid line represents the worldline of a fluid element
passing from region 1 to region 2, the thin solid and arrowed lines denote
the characteristic curves relative to the front\footnote{We recall
		that the characteristic curves
		can be interpreted as the directions in spacetime 
		along which sonic perturbations are transmitted. In 
		this sense it is possible to define the ``region of 
		determinacy'' of an event ${\cal P}$ as the region of 
		spacetime included between the characteristic curves 
		converging to the point ${\cal P}$.} 
 [${\cal C}_{\pm}^j$ are the forward and the backward characteristics of
the regions ahead of the front ($j=1$) and behind it ($j=2$)], the dotted
line shows the the sound speed in the frame of the front and the dashed
line represents the worldline of the front. The characteristic curves are
drawn so to reach the front when the fluid element crosses it and to
depart from the front at the same instant; note that for simplicity we
have assumed the sound speeds to be the same on both sides of the front. 

\vfill\eject \null \vskip -2.75truecm
\centerline{{\psfig{figure=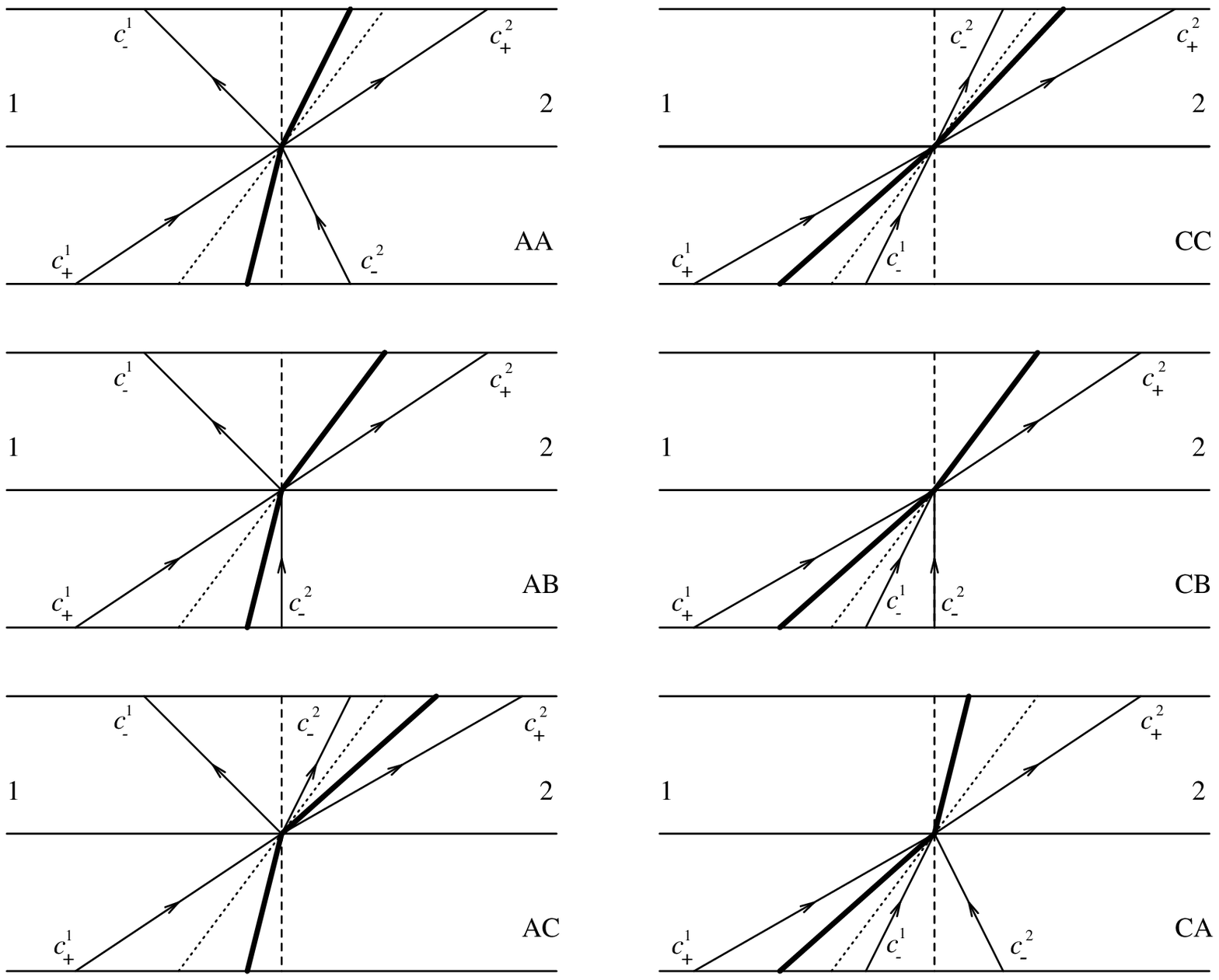,height=13.0truecm,width=15.0truecm}}}
\vskip 1.25truecm \centerline{\vbox{\hsize
14.0truecm\baselineskip=10pt\noindent\small Figure 3. Spacetime diagrams
and characteristic representations of the various types of reaction front
(drawn as left propagating).  Deflagrations are represented on the left
and detonations on the right; the six different diagrams refer
respectively to: AA: weak deflagration, AB: Chapman-Jouguet deflagration,
AC: strong deflagration, CC: weak detonation, CB: Chapman-Jouguet
detonation and CA: strong detonation. The diagrams are drawn relative to
the front rest frame with time on the vertical axis and the space
coordinate on the horizontal one. The thick solid line represents the
worldline of a fluid element passing from region 1 to region 2. For each
diagram, the characteristic curves are denoted with thin solid and arrowed
lines (${\cal C}^j_{\pm}$ , with $j=1,\;2$, are the forward and the
backward characteristics relative to the regions ahead of and behind the
front respectively), the sound speed in the frame of the front is marked
with a dotted line, while the dashed line represents the worldline of the
front. (For simplicity we have assumed the sound speeds to be the same on
both sides of the front.)
}} 

\vfill\eject

\noindent
Deflagrations are represented in the left column, detonations in the right
one and the different diagrams refer respectively to AA: weak
deflagration, AB: Chapman-Jouguet deflagration, AC: strong deflagration,
CC: weak detonation, CB: Chapman-Jouguet detonation and CA: strong
detonation. (The letters A, B, C, express the fact that the fluid element
can move, relative to the front, at a speed smaller, equal or larger than
the sound speed, with the first letter referring to the medium ahead and
the second one to the medium behind.)

	With this representation, the diagrams in Figure 3 show that there
is always a mutual causal connection between the front and the medium
ahead of it when the latter is subsonic (as in the diagrams AA, AB and AC)
and the worldline of the fluid element is within the region of determinacy
of the front. In this case the front can transmit information upstream in
region 1 by means of the backward characteristic ${\cal C}^1_-$. This
mutual causal connection is not present when the medium ahead is
supersonic (as in the diagrams CC, CB and CA) in which case the front
cannot influence the incoming flow and both the characteristic curves
${\cal C}^1_+$ and ${\cal C}^1_-$ are directed towards the front. This
latter property shows that corrugation instabilities of detonations 
fronts cannot propagate in the upstream region 1. 

	Similarly, from Figure 3 it is possible to deduce the causal
connection between the front and the medium behind it, by looking at
whether there is any backward characteristic curve ${\cal C}^2_-$ reaching
the front from region 2. It is evident that this can occur only if the
downstream flow is subsonic (as in diagrams AA and CA) or sonic (as
diagrams AB and CB); this latter case represents a limiting situation,
with the mutual causal connection being just marginal as shown by the
characteristic ${\cal C}^2_-$ being tangent to the worldline of the front.
The front has no mutual causal connection with the medium behind when the
downstream flow is supersonic (as in the diagrams AC and CC) and in this
case the characteristic curves ${\cal C}^2_+$ and ${\cal C}^2_-$ are both
in the direction of the flow. This difference is a fundamental one,
establishing that there is a mutual causal connection between the front
and the medium behind only when the flow out of it is subsonic or sonic.
As a consequence, for subsonic and sonic downstream flows, the medium
behind {\it can} respond to any perturbation produced by the front and, at
least in principle, it can counteract the growth of the potential
instabilities. This backreaction could then appear in the nonlinear
effects mentioned earlier and might be at the origin of the formation of
the typical corrugated but stable cellular flames observed in laboratory
experiments \cite{m51,bl82}, which are produced by weak deflagrations. 

	This stabilizing mechanism {\it cannot} operate in weak
detonations, since in this case there is no mutual interaction between the
front and the media either ahead of it or behind it. We remark that this
argument is simply based on the causal structure of weak detonations and
is therefore independent of the order at which the perturbation analysis
is performed. 

	In the context of causal connection, Chapman-Jouguet
detonations represent a limiting case but do not suffer from the
causality problems discussed for weak detonations. As shown in the
diagram CB of Figure 3 in fact, the front is just marginally mutually 
connected with the medium behind it and this could then provide the
back-reaction required for saturating the potential perturbations
produced at front. 

{\who=1\bigskip\medskip\bigskip\goodbreak
{\Large\bf\noindent\hbox{VII.}\hskip 0.5truecm {Conclusion}
\nobreak \bigskip\medskip\nobreak\who=0}

	We have performed a linear stability analysis of relativistic
strong and Chap-\break man-Jouguet detonations in a relativistic fluid
with specific attention being paid to the stability properties of these
fronts with respect to corrugations. This work has been stimulated by a
recent stability analysis of relativistic Chapman-Jouguet detonations
presented by Abney \cite{a94} with reference to the mechanisms of bubble
growth in first order cosmological phase transitions. In order to
re-examine the above analysis and extend it to strong detonation fronts,
we have implemented the standard linear perturbation techniques on the
hydrodynamical equations for the fluids on either side of the front. The
new perturbed expressions have then been required to be compatible with
the conservation of energy and momentum across a front subject to a
corrugation perturbation. By doing this, we recovered the result that
while no perturbations can be present in the fluid ahead of the front, the
growth of instabilities cannot be ruled out for the medium behind. 

	The study of the time evolution of these potential has been
complicated by the presence of a singular behaviour of the perturbed
hydrodynamical variables if the fluid velocity out of the front is equal
to the local sound speed. This feature, which is common in corrugation
stability studies \cite{ll93,a89} and which did not emerge in the previous
work of Abney \cite{a94}, has been circumvented by using an expansion of
the dispersion relations around the sonic point. 

	As a result of the analysis, we have found that strong detonations
can be both evolutionary and stable with respect to corrugations of the
front, while Chapman-Jouguet detonations appear to be unconditionally
linearly stable fronts. Our conclusions are in contrast with the ones
presented by Abney in \cite{a94} which indicated that Chapman-Jouguet
detonations are effectively unstable at all wavelengths. We believe that
the approximations adopted in \cite{a94} for the derivation of the
perturbed hydrodynamical quantities behind the detonation front are at the
origin of this difference. Because of their intrisic ``underdeterminacy'',
generic weak detonations cannot be studied, but a further equation
accounting for the entropy jump across the front needs to be provided.
This feature unfortunately restricts the stability analysis to specific
situations for which a fourth boundary condition can be expressed and this
will be the subject of a future investigation \cite{r2000}. 

	The results of the stability analysis have been related with the
problem of boundary conditions which emerges when detonation fronts are
used in the scenario of cosmological first order phase transitions such as
the electroweak and quark-hadron transitions. In this context, strong
detonations cannot satisfy the required boundary conditions, despite their
stability properties, and therefore are not suitable for describing the
growth of bubbles of the new phase. Conversely, Chapman-Jouguet
detonations do not suffer restrictions being imposed by the boundary
conditions or by the growth of corrugation instabilities but require a
considerable amount of supercooling of the new temperature phase. Weak
detonations, on the other hand, can satisfy boundary conditions for the
growth of spherical bubbles and could be induced with rather small amounts
of supercooling. However, their causal structure is such that the lack of
a {\it mutual connection} between the front and the media on either side,
would not allow for higher order effects to intervene so as to stabilize
the instabilities that could have possibly been produced.  This feature of
weak detonations represents a strong motivation for the study of their
stability properties and probably hints to the inadequacy of laminar flow
approaches for the study of these fronts. 

\bigskip
\bigskip
{\bigskip\bigskip\medskip\noindent
\Large\bf{\hskip 0.5 truecm Acknowledgments}\bigskip} 

	I would like to thank John Miller for his numerous guiding
comments during the development of this work and P. Haines, O. Pantano and
G. Stoppato for carefully reading this manuscript. I also acknowledge
helpful discussions with M. Abney, M. Laine, B. Link and S. Massaglia.
Financial support for this research has been provided by the Italian
Ministero dell'Universit\`a e della Ricerca Scientifica e Tecnologica. 
Finally, I would like to acknowledge important suggestions of the
anonymous referee that greatly improved the presentation of this work.

\vfill\eject 
\bigskip 
{\bigskip\bigskip\medskip\noindent \Large\bf
{\hskip 0.5truecm Appendix}
\bigskip}

	In this Appendix, details are presented of different expressions
which are discussed in the main text of the paper. The first one concerns
the stability properties of Chapman-Jouguet detonations as they are
deduced from the solutions (\ref{cjw34}) of the dispersion relation
(\ref{drcjf}). The second one is focussed on the presentation of the
complete and generic expressions of the dispersion relation (\ref{dr}) in
the case of strong detonations. 

\vskip 1.0 truecm

	Let us start with considering the form of the roots $\omega_{3,4}$
of (\ref{drcjf}) which can also be written as

\begin{equation}
\label{cjw34n}
\omega_{3,4}=- k_c \left[ i \mp \left( {k^2\over{k^2_c}} - 
1 \right)^{1/2} \right] \;, 
\end{equation}

\noindent
where $k_c$ is a critical wavenumber defined as 

\begin{equation}
k_c = {\gamma^2_2 c_s w_2 (v_1-c_s)(1-v_1c_s) \over{v_1\sigma}} \;. 
\end{equation}

\noindent
Since $k_c$ is a positive real number, the sign of the square root in  
(\ref{cjw34n}) depends on whether the wavenumber for the perturbation 
mode $k$ is larger or smaller than the critical wavenumber. 
The imaginary part of $\omega_{3,4}$ can then be 

\begin{equation}
{\rm Im}\ \omega_{3,4} = -k_c < 0 \,, 
\hskip 4.3 truecm {\rm if}\ k \geq k_c \;, 
\end{equation}
\noindent 
or 

\begin{equation}
{\rm Im}\ \omega_{3,4} = -k_c \left[ 1 \mp 
\left( 1 - {k^2\over{k^2_c}} \right)^{1/2} \right ]\,, 
\hskip 1.5 truecm {\rm if}\ k < k_c \;.  
\end{equation}

\noindent
 It is easy to see that in both cases and irrespective of which of the two
roots is chosen, the imaginary part of the solutions of the dispersion
relation is always negative, thus establishing the stability properties of
of Chapman-Jouguet detonation. 

\bigskip

	Next we turn to the general expression of the dispersion relations
which are deduced after lengthy but straightforward algebraic
manipulations. In the case of strong detonation fronts and writing $v_2 =
c_s+\epsilon$, (with $\epsilon$ being suitably small), equation (\ref{dr})
becomes

\begin{eqnarray}
\label{drsdgen}
\Biggl\{ i\left(1-{c_s+\epsilon\over{v_1}}\right)
\Biggl\{ \left[(1-v_1c_s){2\over{c_s}}-
{(1+v_1c_s)\over{c_s^2}}\epsilon+
{(c_s^4-2c_s^2+1)\over{c_s^3(1-c_s^2)^2}}\epsilon^2 
\right]\omega^3 
\hskip 0.5truecm \nonumber \\
+\; k^2 \Biggl \{ {2 c_s(1-v_1c_s)\over{1-c_s^2}}+
{{[c_s(3+c_s^2)+v_1(2 c_s^4-7 c_s^2+1)]}
\over{c_s(1-c_s^2)^2}}\epsilon
\hskip 2.5truecm \nonumber \\
+{{[c_s(c_s^4+14 c_s^2+1)+v_1(c_s^6-26 c_s^4-7 c_s^2-4)]}
\over{2 c_s^2(1-c_s^2)^3}} \epsilon^2 \Biggr \} \omega
\hskip 3.0truecm  \nonumber \\
+ \left[ {v_1c_s\over{1-c_s^2}}\epsilon+
{{[v_1(3+2c_s^2)-c_s]}\over{2(1-c_s^2)^2}}\epsilon^2 \right]
{k^4\over{\omega}}-{v_1 k^6 c_s^2
\over{2(1-c_s^2)^2 \omega^3}}\epsilon^2 \Biggr\}  
\hskip 3.0truecm  \nonumber \\
+{\sigma\over{w_2}}\Biggl\{ 
\left[(1-c_s^4)c_s^2-c_s(2c_s^4+c_s^2+1) \epsilon
+(1-c_s^4)\epsilon^2\right]
{\omega^2\over{c_s^4}}
\hskip 2.0truecm \nonumber \\
+(c_s^2+1)k^2
+{{(2c_s^2+1)k^2}\over{c_s}}\epsilon
+{{(2c_s^2-1)k^2}\over{2c_s^2}}\epsilon^2
-{k^4\over{2(1-c_s^2)\omega^2}}\epsilon^2 
\Biggr\}\Biggr\}{1\over{\epsilon(2c_s+\epsilon)}} = 0 \>, \nonumber \\
\end{eqnarray}
\noindent
which reduces to (\ref{drsd}) when $c_s^2=1/3$. 

\bigskip

	At last, we show that the quantity $(\omega^2+\gamma_2^2 c_s^2
k^2$) in equation (\ref{drcjf}) should always be different from zero (a
similar statement can be found in \cite{hkllm93}). Because in the
Newtonian case the expressions are simpler and can be handled analytically
we will give the proof in this limit of small velocities. However, the
extension of the result to the special relativistic case is rather
straightforward. As discussed in Section III, the dispersion relation can
be obtained once requiring that the perturbed state vector satisfies the
junction conditions for the energy and momentum. In the limit of small
velocities, we can neglect all the $\gamma$ Lorentz factors and the
components of the perturbed state vector can be simply written as [we here
omit a common factor \hbox{${\rm exp}\; [-i(\omega t + k_y)]$} and drop
the indices referring to region 2]

\begin{eqnarray}
\label{sols}
\delta p &=& c_2 \gamma^2 w {(\omega + l v_x) \over {(\omega v_x + l )}} 
e^{-ilx} \simeq c_2 w {(\omega + l v_x) \over {l}} e^{-ilx} 
\nonumber \\
\delta v_x &=& c_1 e^{i (\omega/v_x) x} + c_2 e^{-ilx} 
\nonumber \\
\delta v_y &=& c_1 {\omega \over {v_x k }} e^{i (\omega/v_x) x} +   
c_2 {k \over {(\omega v_x + l )}} e^{-ilx} \simeq
c_1 {\omega \over {v_x k}} e^{i (\omega/v_x) x} + 
c_2 {k \over {l}} e^{-ilx} 
\nonumber \\
\end{eqnarray}

\medskip\noindent
 where $l\equiv {\tilde l}_2$ and we have used ${\tilde l}_1 = - \omega
/v_x$. Note that we have here exploited the possibility for a different
normalization of the eigenvectors $\vec {\bf L}_1,\; \vec {\bf L}_2$ and 
we are not restricting the discussion to the case $x=0$. 

	Imposing the condition (\ref{u=v}) and asking for the determinant
of the coefficients to be zero is equivalent to set

\begin{equation}
\label{ndet=0}
{\rm det}\ \left(
\begin{array}{ccc}
0 & w \displaystyle {{(\omega + l v_x) \over {l}}} e^{-i l x} &
\delta p \\ \\
e^{i (\omega/v_x) x} & e^{-i l x} & 
\delta v_{x} \\ \\   
\displaystyle {{\omega\over {v_x k}}} e^{i (\omega/v_x) x}&
\displaystyle {{k \over{l}}} e^{-i l x} & 
\delta v_{y}
\end{array}
\right) = 0\;. \nonumber \\
\end{equation}

\medskip\noindent
 A solution of (\ref{ndet=0}) is clearly given by 

\begin{equation}
\label{imo}
\omega = -l v_x = i v_x k \;,
\end{equation}

\medskip\noindent
 where we have used, from (\ref{l23}), that $l=l_2\simeq -i k$.  However,
the solution (\ref{imo}), which represents the small velocity limit of the
positive root of (\ref{cjw12}), is a spurious one and should discarded
since it would imply that $c_1+c_2=0$, and the corresponding solution
(\ref{sols}) would be then identically zero.

\vfill\eject

\end{document}